\documentclass[aps,prl,showpacs,twocolumn,floats,epsfig,pdflatex]{revtex4}
\usepackage{amssymb}
\usepackage{amsbsy}
\usepackage{amsmath}
\usepackage{epsfig}
\usepackage{graphicx}

\begin{document}

\title {Non-equilibrium dynamics of the Bose-Hubbard model: A projection operator approach}

\author{C. Trefzger$^{(1)}$ and K. Sengupta$^{(2)}$}

\affiliation{$^{(1)}$ ICFO - Institut de Ciencies Fotoniques,
Mediterranean Technology Park,
08860 Castelldefels (Barcelona), Spain.\\
$^{(2)}$Theoretical Physics Department, Indian Association for the
Cultivation of Science, Jadavpur, Kolkata-700032, India. }

\date{\today}

\begin{abstract}

We study the phase diagram and non-equilibrium dynamics, both
subsequent to a sudden quench of the hopping amplitude $J$ and
during a ramp $J(t)=Jt/\tau$ with ramp time $\tau$, of the
Bose-Hubbard model at zero temperature using a projection operator
formalism which allows us to incorporate the effects of quantum
fluctuations beyond mean-field approximations in the strong coupling
regime. Our formalism yields a phase diagram which provides a near
exact match with quantum Monte Carlo results in three dimensions. We
also compute the residual energy $Q$, the superfluid order parameter
$\Delta(t)$, the equal-time order parameter correlation function
$C(t)$, and the wavefunction overlap $F$ which yields the defect
formation probability $P$ during non-equilibrium dynamics of the
model. We find that $Q$, $F$, and $P$ do not exhibit the expected
universal scaling. We explain this absence of universality and show
that our results compare well with recent experiments.

\end{abstract}

\pacs{64.60.Ht, 05.30.Jp, 05.30.Rt}

\maketitle

Ultracold atoms in optical lattices provide us with an unique setup
to study non-equilibrium quantum dynamics of closed quantum systems
\cite{bloch1,exp1}. The theoretical study of such quantum dynamics
has seen great progress in recent years \cite{rev1}. Most of these
theoretical works have either restricted themselves to the physics
of integrable and/or one-dimensional (1D) models or concentrated on
generic scaling behavior of physical observables for sudden or slow
dynamics through a quantum critical point (QCP) \cite{rev1}.
However, quantum dynamics of specific experimentally realizable
non-integrable models in higher spatial dimension $d$ and strong
coupling regime have not been studied extensively. The Bose-Hubbard
model with on-site interaction strength $U$ and nearest neighbor
hopping amplitude $J$, which provides an accurate description for
ultracold bosons in an optical lattice, constitutes an example of
such models \cite{bosepapers1}. Most of the studies on dynamics of
this model have concentrated on numerics for $d \le 2$
\cite{koll1,anatoly1}, weak coupling regime \cite{gppapers1}, and
mean-field description of quench dynamics in the strong coupling
regime \cite{ehud1,others1}. Recent experiments \cite{exp1} on
higher dimensional Bose-Hubbard models in the strong-coupling regime
($U \gg J$) clearly necessitate computation of dynamical evolution
of several quantities beyond the mean-field theory and for arbitrary
ramp time $\tau$. To the best of our knowledge, such a study has not
been carried out.

In this work we present a theoretical formalism beyond mean-field
theory, which enables us to investigate in a semi-analytic way, at
equal footing, both the equilibrium phase diagram and the
non-equilibrium dynamics of the Bose-Hubbard model in the strong
coupling regime and at zero temperature. The central results of our
work are the following. First, we compute the equilibrium phase
diagram and demonstrate that it provides a near perfect match to the
corresponding quantum Monte Carlo (QMC) results \cite{qmc1} in three
dimensions (3D). Second, we apply our formalism to non-equilibrium
dynamics of the model for a finite ramp $J(t)=Jt/\tau$ from $t=t_i$
to $t=t_f$. We compute the residual energy $Q$ and the wavefunction
overlap $F$ [{\it i.e.} the overlap between the system wavefunction
after the ramp and the corresponding ground state wavefunction with
$J=J(t_f)$] which also yields the defect formation probability
$P=1-F$ \cite{rev1} as functions of $\tau$. We show that for slow
ramps $P$ reaches a plateau, showing absence of expected scaling
behavior \cite{rev1}. We qualitatively explain such an absence of
universal scaling and relate it to the recent experimental
observations of Ref.\ \cite{exp1}. Finally, we show that our
formalism allows us to address the time evolution of the bosons
after a sudden quench from the Mott ($J=J_i$) to the superfluid
($J=J_f$) phase through the tip of the Mott lobe. We compute the
order parameter $\Delta(t)$ and the equal-time order parameter
correlation function $C(t)$ during such an evolution. We also
compute $F$ and $Q$ for a sudden quench from the critical point
($J_i=J_c$) to the superfluid phase and show that they agree to the
finite ramp results in the limit of small $\tau$ and do not exhibit
universal scaling behavior \cite{anatoly2}. We note that dynamical
properties of the Bose-Hubbard model for $d>2$ in the strongly
coupled regime have not been addressed beyond mean-field theory so
far; our semi-analytical results therefore constitute significant
extension of our understanding of the dynamics of this model in the
strong-coupling regime.

The Hamiltonian of the Bose-Hubbard model is
\begin{eqnarray}
{\mathcal H} &=& \sum_{\langle {\bf r}{\bf r'}\rangle} -J b_{{\bf
r}}^{\dagger} b_{{\bf r'}} + \sum_{{\bf r}} [-\mu {\hat n}_{{\bf r}}
+ \frac{U}{2} {\hat n}_{{\bf r}}({\hat n}_{{\bf r}}-1) ],
\label{ham1}
\end{eqnarray}
where $b$ (${\hat n}$) is the boson annihilation (number) operator
living on the sites of a $d$-dimensional hypercubic lattice, and
the chemical potential $\mu$ fixes the total number of particles.
The corresponding many-body
Schr\"odinger equation $i \hbar
\partial_t |\psi\rangle = {\mathcal H} |\psi \rangle$ is difficult
to handle even numerically due to the infinite dimensionality of the
Hilbert space. A typical practice is to use the Gutzwiller ansatz
$|\psi \rangle = \prod_{{\bf r}} \sum_n c_{n}^{({\bf r})} |n\rangle$
and solve for $c_n^{({\bf r})}$ keeping a finite number of states
$n$ around the Mott occupation number $n=\bar{n}$. This yields the
standard mean-field results with $c_{n}^{({\bf r})}=c_n$ for
homogeneous phases of the model \cite{guzpap1}.

\begin{figure}
\rotatebox{0}{\includegraphics*[width=0.8\linewidth]{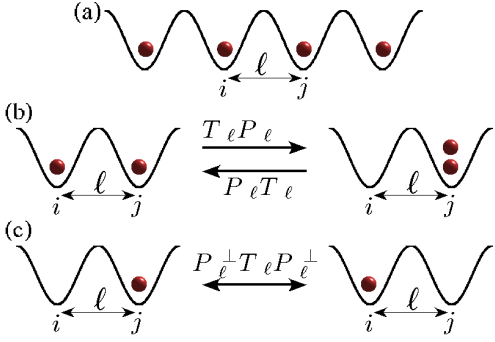}}
\caption{(Color online) (a) Schematic representation of the Mott
state with $\bar n=1$. (b) Typical hopping process mediated via
$T_{\ell}^0$. (c) Hopping process mediated via $P_{\ell}^{\perp}
T_{\ell} P_{\ell}^{\perp}$. Notice that the states in (c) become
part of the low-energy manifold near the critical point, while those
in the right side of (b) do not and are always at an energy $U$
above the Mott state.} \label{figz}
\end{figure}

To build in fluctuations over such a mean-field theory, we use a
projection operator technique \cite{issac1}. The key idea behind
this approach is to introduce a projection operator $P_{\ell}=
|\bar{n}\rangle \langle \bar{n}|_{\bf r} \times |\bar{n}\rangle
\langle \bar{n}|_{\bf r'}$, which lives on the link $\ell$ between
the two neighboring sites ${\bf r}$ and ${\bf r'}$. The effect of
$P_{\ell}$ is to project any state of the system to the manifold of
states for which $n_{\bf r},n_{\bf r'} = \bar n$. Using $P_{\ell}$,
one can rewrite the hopping term of ${\mathcal H}$: $T'
=\sum_{\langle {\bf r} {\bf r'} \rangle} -J b_{{\bf r}}^{\dagger}
b_{\bf r'} = \sum_{\ell} T_{\ell} = \sum_{\ell} [ (P_{\ell} T_{\ell}
+ T_{\ell} P_{\ell}) + P_{\ell}^{\perp} T_{\ell} P_{\ell}^{\perp}]$,
where $P_{\ell}^{\perp}=(1-P_{\ell})$. Note that, as schematically
explained in Fig.\ \ref{figz}, in the strong-coupling regime, the
term $T_{\ell}^0[J] = (P_{\ell} T_{\ell} +T_{\ell} P_{\ell})$
represents hopping processes which take the system out of the
low-energy manifold \cite{issac1}. To obtain an effective low energy
Hamiltonian, we therefore devise a canonical transformation via an
operator $S \equiv S[J]= \sum_{\ell} i[P_{\ell},T_{\ell}]/U$, which
eliminates $T_{\ell}^0[J]$ to first order in $z_0J/U$, where
$z_0=2d$ is the coordination number of the lattice. This leads to
the effective Hamiltonian $H^{\ast}=\exp(iS) {\mathcal H} \exp(-iS)$
up to ${\rm O}(z_0^2 J^2/U)$
\begin{eqnarray}
H^{\ast} &=& H_0 + \sum_{\ell} P_\ell^\perp T_\ell P_\ell^\perp -
\frac{1}{U} \sum_{\ell} \Big[P_\ell T_{\ell}^2 +
 T_{\ell}^2 P_\ell \nonumber\\
&-& P_{\ell} T_{\ell}^2 P_{\ell} - T_{\ell} P_{\ell} T_\ell \Big] -
\frac{1}{U} \sum_{\langle \ell \ell^\prime \rangle} \Big[ P_{\ell}
T_\ell T_{\ell^\prime} - T_\ell P_\ell T_{\ell^\prime} \nonumber\\
&+& \frac{1}{2}\Big ( T_\ell P_{\ell} P_{\ell^\prime}
T_{\ell^\prime} - P_{\ell} T_\ell T_{\ell^\prime} P_{\ell^\prime}
\Big)  + {\rm h.c.} \Big], \label{ham2}
\end{eqnarray}
where $H_0$ denotes the on-site terms in Eq.\ \ref{ham1}. Using
$H^{\ast}$ one can now compute the ground state energy $E= \langle
\psi|{\mathcal H} | \psi \rangle = \langle \psi'|H^{\ast}|
\psi'\rangle + {\rm O}(z_0^3 J^3/U^2)$, where $|\psi'\rangle =
\exp(iS) |\psi\rangle$. We use a Gutzwiller ansatz
\begin{equation}
|\psi'\rangle = \prod_{\bf r} \sum_{n} f_n^{({\bf r})} |n\rangle,
\label{gzwav}
\end{equation}
so that $|\psi'\rangle= |\psi\rangle$ only in the Mott limit
($S,J=0$) and the energy becomes a functional of the coefficients
\begin{eqnarray}
E[\{f_n\};J]= \langle \psi'|H^{\ast}| \psi'\rangle. \label{en1}
\end{eqnarray}
\begin{figure}
\rotatebox{0}{\includegraphics*[width=\linewidth]{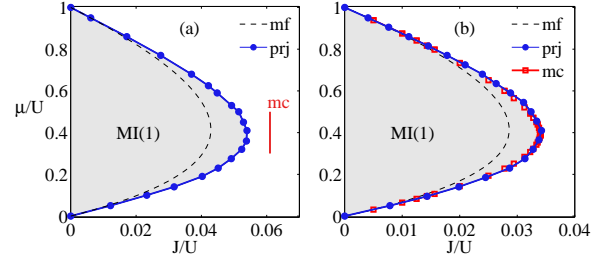}}
\caption{(Color online) Phase diagram of the Bose-Hubbard model in
2D (a) and 3D (b). The blue dots and blue solid lines (black dashed
line) indicate the phase diagram obtained by the projection operator
(mean-field) method. The red squares indicate QMC data.}
\label{fig1}
\end{figure}
The contributions to $E[\{f_n\};J]$ from the first two terms in
$H^{\ast}$ (Eq.\ \ref{ham2}) represent the mean-field energy
functional, while the rest of the terms yield contributions due to
quantum fluctuations. Thus the method constitutes systematic
inclusion of effects of quantum fluctuation over mean-field theory.
The phase diagram obtained by minimizing $E[\{f_n\};J]$ with respect
to $\{f_n\}$ for 2D(3D) and $\bar{n}=1$ is shown in Fig.\
\ref{fig1}(a)(Fig.\ \ref{fig1}(b)). We find that the match with QMC
data \cite{qmc1} is nearly perfect in 3D (with an error of $\sim
0.05 \%$ at the tip of the Mott lobe) where mean-field theory
provides an accurate starting point. In contrast, for 2D, we find
$J_c/U=0.055$ compared to the QMC value $0.061$ (red line in Fig.\
\ref{fig1}(a)). Here the match with QMC is not as accurate; however
it compares favorably to other analytical methods \cite{fr1}. For
the rest of this work, we shall restrict ourselves to $d=3$ and
$\bar{n}=1$.

Next, we apply our formalism to address the dynamics of the model
during a ramp with finite rate $\tau^{-1}$. We consider a ramp
process under which $J$ evolves from $J_i$ at $t_i=0$ to $J_f$ at
$t_f =\tau$:$J(t) = J_i + (J_f-J_i)t/\tau$. To solve the
Schr\"odinger equation, we make a time-dependent canonical
transformation via a time-dependent $S[J(t)]$ to eliminate
$T_{\ell}^0$ up to first order from ${\mathcal H}$ {\it at each
instant}. This yields the Schr\"odinger equation
\begin{eqnarray}
(i \hbar \partial_t +
\partial S/\partial t)|\psi'\rangle = H^{\ast}[J(t)]|\psi'\rangle.
\label{eqqham1}
\end{eqnarray}
The additional term $\partial S/\partial t$ takes into account the
possibility of creation of excitations during the time evolution
with a finite ramp rate $\tau^{-1}$. The above equation yields an
accurate description of the ramp with $H^{\ast}[J(t)]$ given by Eq.\
\ref{ham2} for $J(t)/U\ll 1$. Note that this does not impose a
constraint on $\tau$; it only restricts $J_f/U$ and $J_i/U$, to be
small. Thus the method can treat both "slow" and "fast" ramps at
equal footing.

Substituting Eq.\ \ref{gzwav} into Eq.\ \ref{eqqham1} and allowing
for time-dependent $f_n^{({\bf r})}$, we find that the evolution of
the system is given by the set of coupled equations
\begin{eqnarray}
i \hbar \partial_t f_n^{({\bf r})} &=& \delta E[\{f_n(t)\};J(t)]/
\delta f_n^{\ast ({\bf r})} + \frac{i\hbar}{U} \frac{\partial J(t)}{\partial t}\label{feq1} \\
&\times& \sum_{\langle  {\bf r'} \rangle_{\bf r}} \sqrt{n}
f_{n-1}^{({\bf r})}
\Big[\delta_{n\bar{n}}\varphi_{{\bf r'}\bar{n}} - \delta_{n,\bar{n}+1}\varphi_{{\bf r'},\bar{n}-1}\Big] \nonumber \\
&& + \sqrt{n+1} f_{n+1}^{({\bf r})} \Big[
\delta_{n\bar{n}}\varphi_{{\bf r'},\bar{n}-1}^* -
\delta_{n,\bar{n}-1} \varphi_{{\bf r'}\bar{n}}^* \Big], \nonumber
\end{eqnarray}
where $\varphi_{\bf r} = \langle{\psi^\prime}| b_{\bf r}|
{\psi^\prime}\rangle = \sum_n \varphi_{{\bf r}n} = \sum_n \sqrt{n+1}
f_\mathrm{n}^{*({\bf r})} f_\mathrm{n+1}^{({\bf r})}$, and
$\delta_{n n^\prime}$ is the Kronecker delta.

\begin{figure}
\rotatebox{0}{\includegraphics*[width= 0.9 \linewidth]{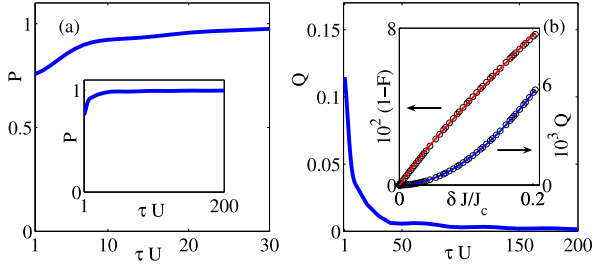}}
\caption{(Color online) (a) Plot of $P$ as a function $\tau U$ (in
units of $\hbar=1$) for $J_i/U=0.05$ (SF phase) and $J_f/U=0.005$
(Mott phase) showing the plateau-like behavior at large $\tau$, and
the corresponding saturation of $Q$ (b). The inset in (b) shows $Q$
and $1-F$ as a function of $\delta J/J_c$ for $\tau U=1$. }
\label{fig4}
\end{figure}

Using Eq.\ \ref{feq1}, we solve for $f_n^{({\bf r})}\equiv f_n$ for
translationally invariant systems numerically keeping all states
$0\le n\le 5$ with $\bar n=1$. Using these, we compute the defect
formation probability $P= 1-F= 1-|\langle
\psi_G|\psi(t_f)\rangle|^2$, where $|\psi_G\rangle$
($|\psi(t_f)\rangle$) denotes the final ground state (state after
the ramp), for a ramp from $J_i/U=0.05$ (superfluid phase) to
$J_f/U=0.005$ (Mott phase) as a function of $\tau$. We find that $P$
exhibits a plateau like behavior at large $\tau$ and do not display
universal scaling as expected from generic theories of slow dynamics
of quantum systems near critical point \cite{rev1}. This seems to be
in qualitative agreement with the recent experiments presented in
Ref.\ \cite{exp1}, where ramp dynamics of ultracold bosons from
superfluid to the Mott region has been experimentally studied.
Indeed, it was found, via direct measurement of ${\bar n}$ per site,
that $P$ displays a plateau like behavior similar to Fig.\
\ref{fig4}(a) [the inset displays the saturation for longer $\tau$].
In Fig.\ \ref{fig4}(b), we show the analogous saturation and lack of
universal scaling of the residual energy $Q=\langle
\psi_f|\mathcal{H}[J_f]|\psi_f\rangle -E_G[J_f]$, where $E_G[J_f]$
denotes the ground state energy at $J=J_f$ as obtained by minimizing
$E[\{f_n\};J_f]$ in Eq.\ \ref{en1}.

Such a lack of universality in the dynamics can be qualitatively
understood from the absence of contribution of the critical (${\bf
k}=0$) modes. In the strong-coupling regime ($J/U \ll 1$), the
system can access the ${\bf k}=0$ modes after a time ${\mathcal T}$,
which can be roughly estimated as the time taken by a boson to cover
the linear system dimension $L$. For typical small $J$ ($U=1$) in
the Mott phase and near the QCP, ${\mathcal T} \sim {\rm
O}(L\hbar/J)$ can be very large. Thus for $t\le{\mathcal T}$, the
dynamics, governed by local physics, which is well captured by our
method, do not display critical scaling behavior. We note that in
realistic experimental setups in the deep Mott limit \cite{exp1},
${\mathcal T}$ may easily exceed the system lifetime making
observation of universal scaling behavior impossible in such setups.

We now apply this method to address the dynamics of the model after
a sudden quench \cite{comment02}, from $J_i$ (Mott phase) to $J_f$
(superfluid phase) through the tip of the Mott lobe, where the
dynamical critical exponent $z=1$. The time evolution of the order
parameter $\Delta_{\bf r}(t) = \langle \psi(t)|b_{\bf
r}|\psi(t)\rangle = \langle \psi'(t)|b'_{\bf r}|\psi'(t)\rangle$,
where $b_{\bf r}'= \exp(iS[J_f]) b_{\bf r} \exp(-iS[J_f])$, can then
be expressed in terms of $f_n^{({\bf r})}$ as

\begin{figure}
\rotatebox{0}{\includegraphics*[width=\linewidth]{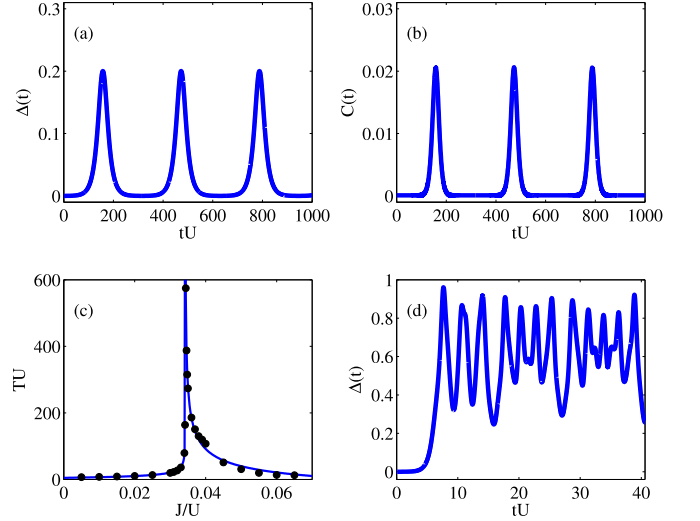}}
\caption{(Color online) Plot of $\Delta(t)$ (a) and $C(t)$ (b) as a
function of $tU$, for $J_f=1.02J_c$.  (c) The time period $T$ of the
oscillations of $\Delta(t)$. (d) Same as in (a) for $J_f=3.51J_c$.
We have set $\hbar=1$ for all plots.} \label{fig2}
\end{figure}

\begin{eqnarray}
\Delta_{\bf r}(t) &=& \varphi_{\bf r}(t) + J/U \sum_{\langle {\bf
r'} \rangle_{\bf r}} \bar{n}
\Big[|f_{\bar{n}}^{({\bf r})}|^2 - f_{\bar{n}-1}^{({\bf r})}|^2 \Big] \varphi_{{\bf r'}\bar{n}} \nonumber \\
&+& (\bar{n}+1) \Big[|f_{\bar{n}}^{({\bf r})}|^2 - f_{\bar{n}+1}^{({\bf r})}|^2 \Big]  \varphi_{{\bf r'},\bar{n}-1}
+ \Big[\Phi_{{\bf r},\bar{n}-2} \nonumber \\
&-&  \Phi_{{\bf r},\bar{n}-1} \Big] \varphi_{{\bf r'}\bar{n}}^* +
\Big[\Phi_{{\bf r}\bar{n}} - \Phi_{{\bf r},\bar{n}-1} \Big]
\varphi_{{\bf r'},\bar{n}-1}^*,  \label{odyn1}
\end{eqnarray}
where $\Phi_{{\bf r} n} = \sqrt{(n+1)(n+2)} f_\mathrm{n}^{*({\bf
r})} f_\mathrm{n+2}^{({\bf r})}$. Note that the first term in Eq.\
\ref{odyn1} represents the mean-field result. The role of quantum
fluctuations in the evolution of $\Delta_{\bf r}(t)$ becomes evident
in computing the equal-time order parameter correlation function
$C_{\bf r}(t) = \langle \psi'(t)|b'_{\bf r} b'_{\bf r}
|\psi'(t)\rangle -\Delta_{\bf r}^2(t)$. To compute $\Delta_{\bf
r}(t)$ and $C_{\bf r}(t)$, we solve Eq. \ref{feq1} numerically for a
translationally invariant system. The resultant plot of $\Delta_{\bf
r}(t) \equiv \Delta(t)$ is shown in Fig.\ \ref{fig2}(a)[(d)] for
$J_i=0$ and $J_f/J_c=1.02$($J_f/J_c=3.51$). We find that near the
critical point, $\Delta (t)$ displays oscillations with a single
characteristic frequency \cite{ehud1}, while away from the critical
point ($J_f/J_c=3.51$), multiple frequencies are involved in its
dynamics. The time period $T$ (Fig.\ \ref{fig2}(c)) of these
oscillations near $J_c$ is found, as a consequence of critical
slowing down, to have a divergence $T \sim |J_f - J_i|^{-z\nu} =
\delta J^{-0.35\pm 0.05}$, leading to $z\nu = 0.35\pm 0.05$ for
$d=3$ \cite{rev1}. Finally, in Fig.\ \ref{fig2}(b) we plot $C_{\bf
r}(t) \equiv C(t)$ as a function of $t$, for $J_f=1.02J_c$. We find
that $|C(t)/\Delta^2(t)|$ may be as large as $0.5$ at the tip of the
peaks of $\Delta(t)$, which shows strong quantum fluctuations near
the QCP.

\begin{figure}
\rotatebox{0}{\includegraphics*[width=0.8 \linewidth]{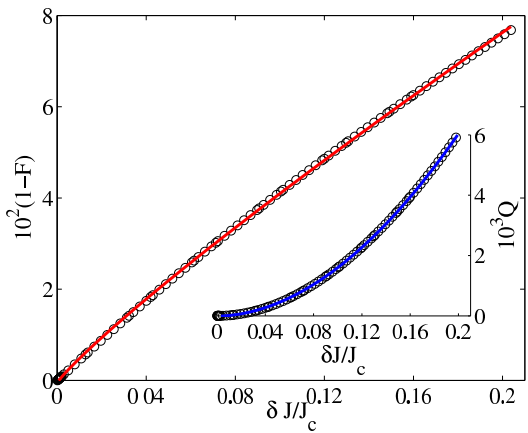}}
\caption{(Color online) Plot of $1-F$ and $Q$ as a function of the
$\delta J$ for $\delta J/J_c \ll 1$. The lines correspond to fits
yielding a power $1-F (Q) \sim (\delta J)^{r_1 (r_2)}$ with $r_1
\simeq 0.89$ and $r_2 \simeq 1.9$.} \label{fig3}
\end{figure}

Next, we compute the wavefunction overlap $F= |\langle
\psi_f|\psi_c\rangle|^2 = |\langle \psi'_f|e^{iS[J_f]}e^{-i S[J_c]}
|\psi'_c\rangle|^2$ for a sudden quench starting at the QCP. Here
$|\psi_f\rangle$($|\psi_c\rangle$) denotes the ground state
wavefunction for $J=J_f(J_c)$. We also compute the residual energy
$Q= \langle \psi_c|\mathcal{H}[J_f]|\psi_c\rangle -E_G[J_f]$, and in
Fig.\ \ref{fig3} we plot $1-F$ and $Q$ for the homogeneous case as a
function of $\delta J = |J_f-J_c|$, for $\delta J/J_c \lesssim 0.2$.
A numerical fit of these curves yields $1-F \sim \delta J^{0.89}$,
and $Q\sim \delta J^{1.90}$, which disagrees with the universal
scaling exponents ($1-F \sim \delta J^{d\nu}$ and $Q\sim \delta
J^{(d+z)\nu}$) expected from sudden dynamics across a QCP with $z=1$
\cite{anatoly2}. Note that our results for the sudden quench match
with those for the ramp dynamics at small $\tau$, shown in the inset
of Fig.\ \ref{fig4}(b). In particular, the exponents obtained from
the two cases are nearly identical, reflecting accurate reproduction
of fast ramp dynamics in the sudden quench limit.

Finally, we estimate the range of physical temperatures for which
the zero temperature theory is accurate. For typical lattice depths
in the Mott or critical regimes, $U \sim 2$ kHz $\simeq 200$nK
\cite{bloch1}. This yields, in 3D, a melting temperature $T^{\ast}
\simeq 0.2 U = 40$nK for the Mott phase and a critical temperature
$T_c \simeq z_0 J_c \simeq 35$nK for the SF phase at the Mott tip
\cite{ketterle01}. This requires the system temperature to be a few
nano-Kelvins (and $ \ll T^{\ast},\, T_c$) , which is well within the
current experimental limit $\sim 1$nK \cite{ketterle01}.

In conclusion, we have presented a projection operator formalism
that describes in a semi-analytical way both the phase diagram, and
non-equilibrium dynamics of the Bose-Hubbard model. It produces a
phase diagram which is nearly identical to the QMC results in 3D,
and allows computation of several quantities such as $F$, $Q$,
$\Delta(t)$, $P$, and $C(t)$ for non-equilibrium dynamics. Its
prediction for $P$ for a slow ramp matches qualitatively with recent
experiments. The method, in principle, can be generalized to
correlated systems which allow perturbative treatment of
fluctuations and for studying ultracold bosons in a finite trap. We
leave such considerations for future study.

The authors thank M. Lewenstein for support and hospitality, E.
Altman, C. Lannert, A. Polkovnikov, and S. Vishveshwara for
discussions, and B. Caprogrosso-Sansone for sharing QMC data. KS
thanks DST, India for support under Project SR/S2/CMP-001/2009. CT
acknowledges support of Spanish MEC (FIS2008-00784, QOIT) and
hospitality of IACS.


\begin{thebibliography}{99}
\bibitem{comment02}
Note that for a sudden quench, $\partial J/\partial t \sim
\delta(t)$, so that one can neglect the $\partial J/\partial t$ term
in Eq.\ \ref{feq1} for addressing dynamical evolution for $t>0$.
\end{thebibliography}
\end{document}